\documentclass[prd,superscriptaddress,twocolumn,preprintnumbers,%
  amsmath,amssymb]{revtex4-1}

\usepackage[breaklinks,colorlinks=true]{hyperref}
\hypersetup{linkcolor=blue, citecolor=cyan}
\usepackage{mathtools}
\usepackage{color}
\usepackage{slashed}
\usepackage{braket}
\allowdisplaybreaks

\begin{document}
\title{Berry phase in the phase space worldline representation: the axial
anomaly and classical kinetic theory}
\author{Patrick Copinger}
\email{copinger0@gate.sinica.edu.tw}
\affiliation{Institute of Physics, Academia Sinica, Taipei 11529, Taiwan}

\author{Shi Pu}
\email{shipu@ustc.edu.cn}
\affiliation{Department of Modern Physics, University of Science and Technology of China, Anhui 230026, China}


\begin{abstract}
The Berry phase is analyzed for Weyl and Dirac fermions in a phase space representation of the worldline formalism. Kinetic theories are constructed for both at a classical level. Whereas the Weyl fermion case reduces in dimension, resembling a theory in quantum mechanics, the Dirac fermion case takes on a manifestly Lorentz covariant form. To achieve a classical kinetic theory for the non-Abelian Dirac fermion Berry phase a spinor construction of Barut and Zanghi is utilized. The axial anomaly is also studied at a quantum level. It is found that under an adiabatic approximation, which is necessary for facilitating a classical kinetic theory, the index of the Dirac operator for massless fermions vanishes. Even so, similarities of an axial rotation to an exact non-covariant Berry phase transform are drawn by application of the Fujikawa method to the Barut and Zanghi spinors on the worldline.
\end{abstract}

\maketitle

\section{Introduction}
\label{sec:intro}

The Berry, or geometric, phase describes the phase accumulation from a nonholonomic motion of a quantum system~\cite{Berry:1984jv},  and successfully explains the Hall current and conductivity~\cite{Thouless:1982zz,*Chang:1995zz,*Chang:1995ebu}, the anomalous Hall effect~\cite{PhysRev.95.1154,*doi:10.1126/science.1058161,*Jungwirth:2002zz,*Fang:2003ir}, and electronic transport properties in condensed matter~\cite{Chang:2008zza,*Xiao:2009rm}, such as for relativistic Fermi-Dirac distributions in Weyl semimetals~\cite{PhysRevB.85.195320,*PhysRevB.88.125427}. The phase is characterized by a topology made visible through the adiabatic theorem~\cite{shapere1989geometric}, and for chiral fermions exist as Weyl nodes in crystal quasi-momentum space~\cite{Chang:2008zza,*Xiao:2009rm} that permit an anomalous inflow governing a chiral (axial) anomaly. It has shown in~\cite{Stephanov:2012ki,Chen:2013iga,Chen:2014cla} that at a classical level--in contrast to it's usual quantum habitat--the effects of an anomaly are present in a Boltzmann equation in what is known as the chiral kinetic theory.

Classical chiral (massless) kinetic theories, (based on an invariant phase space measure due to the Berry phase~\cite{Xiao:2005qw}), have been extended to non-Abelian Dirac fermions~\cite{Chen:2013iga} and gauge fields~\cite{Stone:2013sga}, and to even spatial dimension~\cite{Dwivedi:2013dea}. Also see~\cite{Pu:2017apt} for lattice quantum chromodynamic (QCD) simulations of non-Abelian Berry curvatures. Furthermore, quantum chiral kinetic theories, (nonequilibrium real-time theories exhibiting quantum anomalous phenomena~\footnote{Let us warn that ``quantum'' here is also used by some to denote massive kinetic theories.}), have been well-studied: from Hamiltonian approaches~\cite{Son:2012wh,Son:2012zy}, path integrals~\cite{Stephanov:2012ki,Chen:2013iga,Chen:2014cla,Chen:2015gta},  Wigner function approaches~\cite{Gao:2012ix,*Chen:2012ca,Son:2012zy,Hidaka:2016yjf,*Fang:2022ttm,Huang:2018wdl},  and effective theories~\cite{Son:2012zy,Manuel:2013zaa,Manuel:2014dza,Lin:2018aon,Lin:2019ytz}. Quantum kinetic theories in a worldline setting have also been studied in~\cite{Mueller:2017lzw,*Mueller:2017arw,*Mueller:2019gjj}. Further applications are shown in~\cite{Chen:2016xtg,*Gorbar:2016qfh,*Gorbar:2017toh,*Hidaka:2017auj,*Rybalka:2018uzh,*Dayi:2017xrr,*Ebihara:2017suq,*Hidaka:2018ekt,*Yang:2018lew,*Dayi:2019hod,*Lin:2019fqo,*Lin:2021sjw,*Yang:2021eoz,*Chen:2021azy,*Pu:2014cwa, *Pu:2014fva,*Fukushima:2018osn}. And the numerical simulations for chiral kinetic theories have been developed in the context of heavy-ion collisions in~\cite{Sun:2016nig,*Sun:2016mvh,*Sun:2017xhx,*Sun:2018idn,*Sun:2018bjl,*Zhou:2018rkh,*Zhou:2019jag,*Liu:2019krs}. The chiral radiation transport theory of neutrinos has been recently developed in~\cite{Yamamoto:2020zrs,*Yamamoto:2021hjs} based on a generalized chiral kinetic theory in curved spacetime \cite{Liu:2018xip,*Hayata:2020sqz}.

Likewise, quantum kinetic theories for massive fermions have also been derived based on a quantum field theoretic Wigner function~\cite{Gao:2019znl,*Weickgenannt:2019dks,*Wang:2019moi,Hattori:2019ahi}, with follow-up studies in, e.g.,~\cite{Liu:2020flb,*Guo:2020zpa,*Sheng:2020oqs,*Huang:2020wrr,*Dayi:2020uwx,*Manuel:2021oah,*Wang:2021owk,*Dayi:2021yhf,*Chen:2021rrl}; also, see recent reviews in~\cite{Hidaka:2022dmn,*Gao:2020pfu}. A quantum kinetic theory for massive fermions is also one microscopic approach to describe the spin polarization in the relativistic heavy ion collisions~\cite{STAR:2017ckg,*STAR:2019erd}.  See~\cite{Yi:2021unq, *Yi:2021ryh,*Wu:2022mkr} and references therein for the applications of quantum kinetic theory combined with hydrodynamic simulations to discuss the spin polarization. After integrating over the momentum, the quantum kinetic theory becomes a macroscopic description for spin dynamics, named relativistic spin hydrodynamics~\cite{Hattori:2019lfp,*Fukushima:2020qta,*Fukushima:2020ucl,*Li:2020eon,*She:2021lhe,*Montenegro:2017lvf,*Montenegro:2017rbu,*Florkowski:2017dyn,*Florkowski:2017ruc,*Florkowski:2018fap,*Shi:2020htn,*Shi:2020qrx,*Hongo:2021ona,*Wang:2021ngp,*Wang:2021wqq}. More discussions on the spin polarization in the relativistic heavy ion collisions can be found in the recent reviews~\cite{Wang:2017jpl,*Becattini:2020ngo,*Becattini:2020sww,*Gao:2020vbh,*Liu:2020ymh}. 

While Lorentz covariant kinetic theories exist in a quantum setting, a manifestly covariant classical kinetic theory guided by anomalous effects stemming from an invariant phase space measure is curiously absent~\footnote{Worldline chiral kinetic theories exist~\cite{Mueller:2017lzw,*Mueller:2017arw,*Mueller:2019gjj}, however our construction from an invariant phase space is novel, which also supports our aim of contrasting the anomalous features stemming from both the classical kinetic theory and quantum descriptions.}, and its construction is half of the twofold scope of the current work. Since a Dirac fermion dispersion relation exists in topological insulators~\cite{Qi:2008ew,*PhysRevB.81.159901}, Dirac semimetals~\cite{PhysRevLett.113.027603,*Liu864}, etc. with prominent 3+1 dimensional non-Abelian Berry curvature, the construction of a corresponding classical kinetic theory can prove indispensable for their transport and nonequilibrium study.  A distinct merit of the worldline approach as compared to a perturbative in $\hbar$ Wigner function approach (see, e.g., ~\cite{Gao:2012ix,*Chen:2012ca,Son:2012zy,Hidaka:2016yjf,Huang:2018wdl}), is its non-perturbative construction, which may be important for non-perturbative corrections to kinetic theories that may include for example the Schwinger effect.  Another key merit of the worldline formalism is an all-orders description of the gauge field.  In quantum electrodynamics (QED) and QCD a classical description of a kinetic theory may then serve as a starting point to a nonperturbative in $\hbar$ description of a quantum kinetic theory.  A classical kinetic theory, built on a Lorentz covariant Berry phase, is accessible through a phase space representation of the first-quantized worldline formalism. And equipped with such a Berry phase brings us to the other half of our twofold scope: exploring the relationship of the phase with the axial anomaly.

The Berry phase in a phase space worldline setting provides us with a unique perspective to study the anomaly. One can understand an exact~\footnote{Let us also caution that differing definitions of a Berry phase exist (see~\cite{Fujikawa:2005cn} for a discussion), notably it is common for adiabaticity to be assumed.} Berry phase that includes both diagonal (adiabatic) and off-diagonal parts, on the worldline with the use of a non-Abelian gauge transformation, $U\in G$  for group $G$, of Hamiltonian, $H\in Lie(G)$, under propertime or parametrization, $\tau$:
\begin{equation}
   H\rightarrow U^{-1}HU+i\hbar U^{-1}\frac{d}{d\tau}U \,;
   \label{eq:Hamil_trans}
\end{equation}
this is essentially the well-known transformation in quantum mechanics~\cite{doi:10.1098/rspa.1987.0131}, but with time replaced with propertime. The exact Berry phase is $-i\hbar U^{-1}(dU/d\tau)$. And on the other hand the axial anomaly describes the non-conservation of axial current, broken by quantum effects in a gauge field configuration with non-trivial topology~\cite{Adler:1969gk,*Bell:1969ts}. The connection between the Berry phase and Wess-Zumino terms through a Born-Oppenheimer approximation was first established in~\cite{Stone:1985av}. And there has been ample work comparing and contrasting both the axial anomaly and its potential origins from the Berry phase; see, e.g.,~\cite{shapere1989geometric}. Then in~\cite{Deguchi:2005pc,*Fujikawa:2005cn} a careful study illustrated key differences with application of the adiabatic approximation. We find similarly that while a classical kinetic theory may be built from a worldline phase space Berry phase under adiabaticity, the same adiabaticity criteria employed at the quantum level--we calculate the index for massless fermions~\cite{tHooft:1976snw,*Atiyah:1978ri}--makes the anomaly disappear. 

To achieve a classical kinetic theory in a worldline phase space setting, we make use of a coherent state formalism~\cite{Zhang:1990fy} on the worldline; this process entails an identification of the matrix weighted worldline action and path ordering as a path integral over coherent states.  An intuitive construction built with spinors fortunately exists owing to Barut and Zanghi (BZ)~\cite{PhysRevLett.52.2009,Barut:1988ud,*Barut:1984vnk}.  The formalism has been extended to curved space~\cite{Barut:1987kb}, and can be quantized in propertime in a Schr\"{o}dinger-like picture~\cite{Barut:1985xp,*Barut:1987an}. Coherent states have proved useful in the the non-Abelian Stokes theorem~\cite{Kondo:1998nw,*Kondo:2008su,*Kondo:1999tj}, and the non-Abelian worldline instanton method~\cite{Copinger:2020feb}. Coherent states have also enjoyed application to chiral kinetic theories in several contexts: for non-Abelian Dirac fermions~\cite{Xiao:2005qw} and gauge fields~\cite{Stone:2013sga}, and in a worldline setting with grassmann variables for color and fermionic degrees of freedom~\cite{Mueller:2017lzw,*Mueller:2017arw,*Mueller:2019gjj}. Note that an analogous auxilary formulation may also be utilized~\cite{Barducci:1980xk,*Bastianelli:2013pta,*Ahmadiniaz:2015xoa}.

A phase space representation of the worldline formalism is provided in Sec.~\ref{sec:Phase_space_wordline}. Then a classical kinetic theory for Weyl fermions is derived in Sec.~\ref{sec:Weyl}. Next a Berry phase for Dirac fermions is given in Sec.~\ref{sec:DiracBerry}, along with a BZ coherent state to analyze the phase in Sec.~\ref{sec:BZ_coherent}, and finally a Dirac classical kinetic theory is argued in Sec.~\ref{sec:CKT}. Last, we analyze potential connections to the axial anomaly under a Berry phase both under adiabaticity in Sec.~\ref{sec:vanishing}, and more generally as a gauge transformation using the Fujikawa method in Sec.~\ref{sec:fujikawa}.  

Notations are as follows: We work in Minkowski spacetime with metric $g^{\mu\nu}=\mathrm{diag}(1,-1,-1,-1)$.  For the completely antisymmetric tensor,  $\epsilon^{\mu\nu\alpha\beta}$, we take $\epsilon^{0123}=1$.  Our QED covariant derivative reads $D_{\mu}=\partial_{\mu}+i(e/\hbar c)A_{\mu}$.  The covariant Weyl matrices are $\sigma^\mu=(\mathbb{I}_2,\sigma^i)$ and $\bar{\sigma}^\mu=(\mathbb{I}_2,-\sigma^i)$.  For Dirac matrices we work in the Weyl representation:
\begin{equation}
   \gamma_{0}=\begin{array}{cc} \begin{pmatrix} & \mathbb{I}_2\\ \mathbb{I}_2 \end{pmatrix}\end{array},\;
   \gamma^{i}=\begin{array}{cc}\begin{pmatrix}& \sigma^{i}\\-\sigma^{i}\end{pmatrix}\end{array},\;
   \gamma_{5}=\begin{array}{cc}\begin{pmatrix}-\mathbb{I}_2\\& \mathbb{I}_2\end{pmatrix}\end{array}\,,
\end{equation}
with spin tensor $\sigma_{\mu\nu}=(i/2)[\gamma_{\mu},\gamma_{\nu}]$. And an anti-symmetric tensor in Lorentz indices is indicated with brakets such that $A^{[\mu\nu]}=A^{\mu\nu}-A^{\nu\mu}$. Let us finally remark on some variable specificities: $p^\mu$ and $q^\mu$ represent, respectively, the kinetic and canonical momenta. Subscripts with ``$\textrm{W}$" and ``$\textrm{D}$" refer, respectively, to Weyl and Dirac fermions. Dotted variables denote a total derivative with respect to propertime, e.g., $\dot{x}=dx/d\tau$.

\section{Phase space worldline representation} 
\label{sec:Phase_space_wordline}

Let us begin our discussion of the Berry phase by first reviewing the worldline formalism~\cite{Feynman:1950ir,*Feynman:1951gn,*Schubert:2001he},  but in phase space~\cite{Migdal:1986pz} rather than configuration space.  While the phase space representation enjoyed active study initiated by Midgar's QCD loop calculations~\cite{Migdal:1986pz},  it fell out of favor in place of the configuration space representation.  This is in part due to the analytically tractable quadratic form in the configuration space worldline action; i.e., there is at least an $\mathcal{O}(\dot{x}^{2})$ term present in the action.  Even so, with the aid of the BZ coherent state~\cite{PhysRevLett.52.2009,Barut:1988ud,*Barut:1984vnk} adopted for the worldline, we can demonstrate there are key advantages for certain applications in the usage of a phase space worldline representation. One advantage can be clearly seen in the intuitive extension of a 3 dimensional Weyl Hamiltonian,  i.e.,  $H_{\textrm{W}} = \boldsymbol{p} \cdot \boldsymbol{\sigma}$, 
proposed in~\cite{Stephanov:2012ki, Chen:2013iga, Chen:2014cla}, for development of a classical kinetic theory,  to the quantum mechanical-like 3+1 dimensional worldline Dirac Hamiltonian, 
\begin{equation}
   H_{\textrm{D}}= -\slashed{p} + mc\,,
\end{equation}  
with kinetic momentum $p^\mu$, thus affording us,  upon transformation,  with an intuitive picture of the Berry phase in momentum space.  Also, let us point out that worldline phase space representations have enjoyed application to non-commutative Snyder spaces~\cite{Bonezzi:2012vr,*Franchino-Vinas:2018qyk}.

To illustrate the phase space representation let us derive the propagator for a massive fermion in a background electromagnetic field.  We begin with the formal solution~\cite{doi:10.1142/7305}, where we have
assumed a minimal substitution, 
$\partial_\mu\rightarrow D_\mu = \partial_\mu +i(e/\hbar c)A_\mu$, 
and introduce Schwinger propertime such that
\begin{align}
   G(A,x,y)&=\bra{x}\frac{-\hbar}{i\hbar\slashed{\partial}
   -\frac{e}{c}\slashed{A}-mc+i\epsilon}\ket{y} \nonumber \\
   &=\int_{0}^{\infty}dT\,i\bra{x}e^{-\frac{i}{\hbar}(H_{\textrm{D}}-i\epsilon)T}\ket{y}\,,
\end{align}
where now $H_{\textrm{D}}=-\slashed{q}+(e/c)\slashed{A}(x)+mc$, with canonical momentum $q^\mu$. And a small $\epsilon$ is required for convergence of the propertime integral, which from here out we leave implicit. Here, operators are understood based on context. Let us construct the appropriate path integral via a Legendre transformation~\cite{Schwartz:2014sze}. Then in the Heisenberg representation we construct a path integral through a Legendre transformation using $\mathcal{L}=q_{\mu}(\partial H/\partial q^{\mu})-H$. 
One can find with the use of the Hamiltonian equation, $\dot{x}_{\mu}=-(i/\hbar)[x_{\mu},H]=\gamma_{\mu}$, where we have also used $[q_{\mu},x_{\nu}]=i\hbar\,g_{\mu\nu}$. 
Then for kinetic momentum, $p_{\mu}= q_{\mu}-(e/c)A_{\mu}(x)$, we find for the Lagrangian,
\begin{equation}
   \mathcal{L}=-p_{\mu}\dot{x}^{\mu}-\frac{e}{c}A_{\mu}(x)\dot{x}^{\mu}+\slashed{p}-mc\,.
\end{equation}
Finally, using the identity, 
\begin{equation}
      \bra{x}e^{-\frac{i}{\hbar}HT}\ket{y} = \int_{x(0)=y}^{x(T)=x}\mathcal{D}x\int\frac{\mathcal{D}p}{2\pi\hbar}\mathcal{P}e^{\frac{i}{\hbar}\int_{0}^{T}d\tau\mathcal{{L}}}\,, 
\end{equation}
one can find the accompanying path integral form as, 
\begin{align}
   G(A,x,y)&=i\int_{0}^{\infty}dT 
   \int_y^x \mathcal{D}x 
   \int\frac{\mathcal{D}p}{2\pi\hbar}\,e^{\frac{i}{\hbar}S_{\textrm{A}}}\mathcal{W}_{\textrm{D}}\,,
   \label{eq:Green}\\
   S_{\textrm{A}}&\coloneqq\int_{0}^{T}d\tau[-mc-p_{\mu}\dot{x}^{\mu}
   -\frac{e}{c}A_{\mu}\dot{x}^{\mu}]\,,
   \label{eq:S_A}
\end{align}
with a path ordered (acting on Dirac indices) factor being
\begin{equation}
   \mathcal{W}_{\textrm{D}}\coloneqq\mathcal{P} 
   \exp\Bigl\{\frac{i}{\hbar}\int_{0}^{T}d\tau\,\slashed{p}\Bigr\}\,.
   \label{eq:Wilson_Dirac}
\end{equation}
For a derivation from breaking up the propertime,  $T$,  into infinitesimal segments see~\cite{Fosco:2003rr}.  Let us also mention that an effective action can also be constructed in a similar fashion:
\begin{align}
   \Gamma[A]&=-i\hbar\mathrm{Tr}\ln\bigl[i\hbar \slashed{\partial}
   -\frac{e}{c}\slashed{A}(x)-mc\bigr]  \\
   &=i\hbar\mathrm{tr}\int_{0}^{\infty}\frac{dT}{T}\oint\mathcal{D}x
   \int\frac{\mathcal{D}p}{2\pi\hbar}\,e^{\frac{i}{\hbar}S_{\textrm{A}}}\mathcal{W}_{\textrm{D}}\,,   
   \label{eq:Eff_Action}
\end{align}
with appropriate counterterms.  Here 
$\oint\mathcal{D}x=\int dx'\int \mathcal{D}x$ with periodic path integral endpoints $x(0)=x(T)=x'$.

Let us remark that the functional trace in the effective action in Eq.~\eqref{eq:Eff_Action} may instead be expressed in terms of momentum~\cite{doi:10.1142/7305}.  This amounts a replacement of the path integral measure to $\int \mathcal{D}x\oint \mathcal{D}p/(2\pi\hbar)$, where now $\oint \mathcal{D}p =\int d p' \int \mathcal{D} p$ with periodic path integral endpoints, $p(T) = p(0) = p'$.  The periodicity in momentum is important for the realization of a Berry phase in the worldline phase space construction.

In this section, we have introduced the phase space worldline representation for Dirac fermions,  however it is first instructive to examine the massless Weyl fermion case.  For then we can bridge the connection to quantum mechanical classical kinetic theory case~\cite{Stephanov:2012ki,Chen:2013iga, Chen:2014cla}.

\section{Worldline classical kinetic theory for Weyl fermions}
\label{sec:Weyl}

Due to the chiral anomaly spawning from the fermion determinant under chiral rotation~\cite{Fujikawa:1979ay,*Fujikawa:1980eg}, Eq.~\eqref{eq:Wilson_Dirac} may not be separated into left and right parts.  Also it has been found important to take the small mass limit only for final expectation values for the massive theory~\cite{Copinger:2018ftr}, indeed massless QED and QED in the small mass limit are two different theories~\cite{Fomin:1976am}.  Nevertheless it is of theoretical interest to analyze the case of Weyl fermions as they appear as gapless excitations in semimetals~\cite{PhysRevB.85.195320,*PhysRevB.88.125427},  whose study has supplied a new understanding and observation~\cite{PhysRevLett.113.027603,*Liu864} of the anomaly.  Therefore we postulate the following Lagrangian for left component Weyl fermions in Minkowski spacetime as 
\begin{equation}
    \mathcal{L}_{\textrm{W}}= i\hbar \psi_{\textrm{L}}^{\dagger}\bar{\sigma}_{\mu}D^{\mu}\psi_{\textrm{L}}\,.
\end{equation}
Upon integrating out the fermions we are left with the effective action,  which may be represented in the worldline path integral formalism following similar steps as outlined above, whose form is similar to Eq.~\eqref{eq:Eff_Action} however with the replacement
\begin{equation}
   \mathcal{W}_{\textrm{D}}\rightarrow\mathcal{W}_{\textrm{W}}
   \coloneqq\mathcal{P}\exp\Bigl\{\frac{i}{\hbar}\int_{0}^{T}d\tau\,p_{\mu}\bar{\sigma}^{\mu}\Bigr\}\,,
   \label{eq:Wilson_Weyl}
\end{equation}
and also for the massless fermions, $m\rightarrow 0$.

We may diagonlize the 
$p_\mu\bar{\sigma}^\mu$ 
term by breaking up the path integral,  inserting complete sets of eigenstates, $U(\boldsymbol{p})U^{\dagger}(\boldsymbol{p})=\mathbb{I}_2$,
as illustrated in~\cite{Stephanov:2012ki, Chen:2013iga, Chen:2014cla}. The eigendecomposition of the Weyl worldline Hamiltonian reads 
\begin{equation}
    U^{\dagger}p_{\mu}\bar{\sigma}^{\mu}U=p^{0}\mathbb{I}_2+|\boldsymbol{p}|\sigma_{3}\;,
\end{equation}
where importantly the energy remains diagonal and does not contribute to the Berry phase.  Let us next confine our attention to the worldline effective action, which traces over Eq.~\eqref{eq:Wilson_Weyl}. Then after the transformation one may find for the revised path ordered element:
\begin{equation}
   \mathrm{tr}\mathcal{W}_{\textrm{W}}=\mathrm{tr}\mathcal{P}\exp\Bigl\{\frac{i}{\hbar}
   \int_{0}^{T}d\tau\bigl[p^{0}\mathbb{I}_2+|\boldsymbol{p}|\sigma_{3}
   -\boldsymbol{B}_{\textrm{W}}\cdot\dot{\boldsymbol{p}}\bigr] \Bigr\}\,,
   \label{eq:Wilson_Weyl_trace}
\end{equation}
where Berry's connection and curvature are, respectively,
\begin{equation}
    \boldsymbol{B}_{\textrm{W}}=-i\hbar U^{\dagger}\nabla_{\boldsymbol{p}}U\,,\quad \boldsymbol{S}_{\textrm{W}}=\nabla_{\boldsymbol{p}}\times\boldsymbol{B}_{\textrm{W}}\,,
\end{equation}
and we have made use of the cyclicity of the trace and the fact that $U(\boldsymbol{p}(0))=U(\boldsymbol{p}(T))$. 
To find the unitary transformation let us write 
$\boldsymbol{p}=|\boldsymbol{p}|(\sin\theta_{p}\cos\omega_{p},\sin\theta_{p}\sin\omega_{p},\cos\theta_{p})$, 
then the normalized set of eigenvectors, with $U=(u^{-},u^{+})$, are
\begin{equation}
   u^{-}=\begin{pmatrix}e^{-i\omega_{p}}\cos\tfrac{\theta_{p}}{2}
   \\ \sin\tfrac{\theta_{p}}{2}\end{pmatrix}\,,\quad 
   u^{+}=\begin{pmatrix}-e^{-i\omega_{p}}\sin\tfrac{\theta_{p}}{2}\\
   \cos\tfrac{\theta_{p}}{2} \end{pmatrix}\,.
\end{equation}

For large propertimes, $T$,  such that the adiabatic theorem applies and off-diagonal terms in the Berry phase may be dismissed,  Eq.~\eqref{eq:Wilson_Weyl_trace}, decouples into
\begin{equation}
   \mathrm{tr}\mathcal{W}_{\textrm{W}}\approx\sum_{\pm}\exp\Bigl\{\frac{i}{\hbar}
   \int_{0}^{T}d\tau[p^{0}\pm|\boldsymbol{p}|-\boldsymbol{B}_{W}^{\mp}\cdot\dot{\boldsymbol{p}}]\Bigr\}\,,
   \label{eq:Weyl_adiabatic}
\end{equation}
where $\pm$ sums over the positive and negative energy contributions. And the components of the Berry phase for adiabatic transport for the positive and negative energy particles can be found as 
\begin{equation}
    \boldsymbol{B}_{\textrm{W}}^{\pm}=-i\hbar u^{\pm}\nabla_{\boldsymbol{p}}u^{\pm}\,,\;\; \boldsymbol{S}_{\textrm{W}}^{\pm}=\nabla_{\boldsymbol{p}}\times\boldsymbol{B}_{\textrm{W}}^{\pm}=\mp\hbar \frac{\boldsymbol{p}}{2|\boldsymbol{p}|^{3}}\,.
\end{equation}
To construct a kinetic theory,  let us focus on the positive energy particle; here the worldline action reads,
\begin{equation}
   \mathcal{S}_{\textrm{W}}=\int_{0}^{T}d\tau\bigl[-p_{\mu}\dot{x}^{\mu}-\frac{e}{c}A_{\mu}\dot{x}^{\mu}+p^{0}-|\boldsymbol{p}|-\boldsymbol{B}_{\textrm{W}}^{+}\cdot\dot{\boldsymbol{p}}\bigr]\,.
   \label{eq:Weyl_final_action}
\end{equation}
One may note that after a reparameterization, $\tau\rightarrow T\tau'$, the proper time,  $T$, would act as a Lagrange multiplier sending the single particle action on shell,  i.e., $p^{0}=|\boldsymbol{p}|$. The equations of motion for the worldline action above are
\begin{equation}
   \dot{p}_{\mu}=\frac{e}{c}F_{\mu\nu}\dot{x}^{\nu}\,,\quad
   \dot{\boldsymbol{x}}=\frac{\boldsymbol{p}}{|\boldsymbol{p}|}+\boldsymbol{S}_{\textrm{W}}^{+}\times\dot{\boldsymbol{p}}\,,\quad \dot{x}^{0}=1\,.
   \label{eq:weyl_current_berry}
\end{equation}
which are the same as those for the reduced dimension, (i.e., 3+1 $\rightarrow$ 3), quantum mechanics case found in~\cite{Stephanov:2012ki,Chen:2013iga, Chen:2014cla}. Therefore we can see the Weyl worldline Berry phase is the same as its non-covariant counterpart, and therefore analogous treatments with the quantum mechanical case hold.  One may equally well here define a classical kinetic theory by constructing an invariant measure phase space,  $(1+\boldsymbol{S}_{\textrm{W}}^{\pm}\cdot\boldsymbol{B})d^{3}xd^{3}p/(2\pi)^{3}$ with magnetic field $\boldsymbol{B}$, whose introduction into a phase space distribution,  $\rho_{\textrm{W}}=(1+\boldsymbol{S}_{\textrm{W}}^{\pm}\cdot\boldsymbol{B})f_{\textrm{W}}$, modifies a Liouville equation,  and makes possible classically the axial anomaly,  chiral magnetic effect~\cite{Fukushima:2008xe},  and chiral vortical effect~\cite{Vilenkin:1979ui}.  While a phase space worldline classical kinetic theory is the same for the non-relativistic case for Weyl fermions,  for Dirac fermions the worldline construction displays several unique features.

\section{Berry phase for Dirac fermions}
\label{sec:DiracBerry}

Moving our attention to the full case of Dirac fermions we wish to apply a geometric phase transformation to $\mathrm{tr}\mathcal{W}_{\textrm{D}}$, given in Eq.~\eqref{eq:Wilson_Dirac},  as was accomplished for the Weyl case above.  A natural Lorentz invariant choice for the transformation is through the use of spinors. 
Then one may cast the spinors into an $s\in$ SO$(1,3)$ similarity transform. Following Eq.~\eqref{eq:Hamil_trans}, one can then see $\slashed{p} \rightarrow s^{-1}\slashed{p}s+i\hbar s^{-1}\dot{s}$ with accompanied Berry phase, where $s^{-1}=\gamma^{0}s^{\dagger}\gamma^{0}$. The $s$ may be chosen to take 
\begin{equation}
    s^{-1}\slashed{p}s=p\gamma_{0}\,,
    \label{eq:gamma_0}
\end{equation}
where we make use of the following notations throughout:
\begin{equation}
    p \coloneqq \sqrt{ p^\mu p_\mu}\,,\quad \hat{p}_{\mu}=p_{\mu}/p\,.
\end{equation}
A key distinction from Weyl fermion case above and the Dirac Hamiltonian of quantum mechanics, i.e., $\gamma_0\boldsymbol{\gamma}\cdot \boldsymbol{p} + \gamma_0 m$, is that the transformed element of the phase space worldline Dirac fermion, Eq.~\eqref{eq:gamma_0} is proportional to $p$. However, like the Weyl fermion case (for $|\boldsymbol{p}|=0$) discussed above the degenerate point at $p=0$ here is crossed, signaling anomalous effects in the breakdown of adiabaticity~\cite{Stephanov:2012ki}. 

In the Dirac representation of the gamma matrices Eq.~\eqref{eq:gamma_0} would be diagonal,  however we elect to use the Weyl representation.  A hallmark of using Weyl spinors is that $s$ is block diagonal, so therefore so too is the Berry phase, and thus the Berry phase in the Weyl representation can only mix left and right parts.

Let us digress on the generalization of the diagonalization of the worldline phase space Hamiltonian. One may equally well select $s$ such that the transformed element is $p\gamma_5$, which is diagonal in the Weyl representation. And of course one may alternatively use the Dirac representation.  As expected these choices are irrelevant to quantum observables; we will revisit the ambiguity in Sec.~\ref{sec:vanishing}. Let us also note that we will treat a non-covariant eigendecomposition, which does not transform under SO$(1,3)$, in Sec.~\ref{sec:fujikawa}; it yields $p\gamma_5$ and has interesting properties such a covariant Berry phase, in contrast here where a side-jump is present after Lorentz boost~\cite{Chen:2014cla,Hidaka:2016yjf}. 

The explicit eigendecomposition of $\slashed{p}$ is motivated from positive and negative energy eigenspinor solutions such that $\slashed{p}u_{i}=pu_{i}$ and $\slashed{p}v_{i}=-pv_{i}$, with 
$\bar{u}_i=u_i\gamma_0$ and $\bar{v}_i=v_i\gamma_0$.
The set of eigenspinors are orthonormalized so that 
$\bar{u}_{i}u_{j}=-\bar{v}_{i}v_{j}=\delta_{ij}$,
$\bar{u}_{i}v_{j}=0$, and $\sum_{i}u_{i}\bar{u}_{i}-v_{i}\bar{v}_{i}=\mathbb{I}_4$.
A set of solutions in the Weyl gamma matrix representation can be found as~\cite{Peskin:1995ev}
\begin{equation}
   u_{i}=\frac{1}{\sqrt{2}}\begin{array}{c}
   \begin{pmatrix}\sqrt{\hat{p}_{\mu}\sigma^{\mu}}\xi_{i}\\ 
   \sqrt{\hat{p}_{\mu}\bar{\sigma}^{\mu}}\xi_{i}
   \end{pmatrix}\end{array}\,, \;
   v_{i}=\frac{1}{\sqrt{2}}\begin{array}{c}\begin{pmatrix}
   \sqrt{\hat{p}_{\mu}\sigma^{\mu}}\eta_{i}\\
   -\sqrt{\hat{p}_{\mu}\bar{\sigma}^{\mu}}\eta_{i}
   \end{pmatrix}\end{array}\,,
\end{equation}
for the sets of two component spinors, $\xi_{i}$ and $\eta_{i}$.  However,  for simplicity we select 
$\xi_{1}=\eta_{1}=[1,0]^{T}$ and
$\xi_{2}=\eta_{2}=[0,1]^{T}$. The arguments of the square roots are understood to be
\begin{align}
   \sqrt{\hat{p}_{\mu}\sigma^{\mu}}&= (2p(p+p_0))^{-1/2}[p_{\mu}\sigma^{\mu}+p\,\mathbb{I}_2]
   \coloneqq\hat{p}_{\frac{1}{2}}^{\mu}\sigma_{\mu}\\
   \sqrt{\hat{p}_{\mu}\bar{\sigma}^{\mu}}&=(2p(p+p_0))^{-1/2}
   [p_{\mu}\bar{\sigma}^{\mu}+p\,\mathbb{I}_2] = \hat{p}_{\frac{1}{2}}^{\mu}\bar{\sigma}_{\mu}
\end{align}

Gathering the spinors into a similarity transform of the type depicted above one can construct 
$s=(1/\sqrt{2})[u_{1},u_{2},v_{1},v_{2}](\gamma_{0}-\gamma_{5})$, or 
\begin{equation}
   s=\gamma_{\mu}\gamma_{0}\hat{p}_{\frac{1}{2}}^{\mu}\,,\quad 
   s^{-1}=\gamma_{0}\gamma_{\mu}\hat{p}_{\frac{1}{2}}^{\mu}\,,
   \label{eq:sim_transform}
\end{equation}
which takes 
$\slashed{p}s=s\gamma_{0}p$ or 
$s^{-1}\slashed{p}=p\gamma_{0}s^{-1}$.
An exact (containing both adiabatic components and parts that do not commute with the adiabatic components) Berry phase that is a pure gauge transformation of $\slashed{p}$ can be found as
\begin{align}
   B_{\textrm{PG}\mu}&=-i\hbar s^{-1}(\partial_{\mu}^{p}s) \nonumber \\
   &=\frac{\hbar}{2}\frac{1}{p_{0}+p}\gamma_{0}
   \bigl\{(\delta_{\mu}^{\;\nu}-\hat{p}_{\mu}\hat{p}^{\nu})\sigma_{\nu0}
   -\hat{p}^{\nu}\sigma_{\nu\mu}\bigr\}\gamma_{0}\,.
    \label{eq:pure_gauge}
\end{align}

Let us now remark on the adiabatic theorem as it applies to the worldline construction with propertime taking the place of real time. It is sufficient to examine solely the path ordered expression in Eq.~\eqref{eq:Wilson_Dirac} that appears in the path integral.  Then in analogy to the the quantum mechanical case, we can recognize an adiabaticity for both large propertimes, $T$,  and large gap separating eigenvalues of the positive and negative modes, $2p$: $2pT\gg2\pi \hbar$; see, e.g.,~\cite{Stephanov:2012ki, Chen:2013iga, Chen:2014cla}.  Let us write this step as
\begin{align}
   \lim_{\textrm{Ad}}\mathcal{W}_{\textrm{D}} 
  & \coloneqq \lim_{pT\gg\pi\hbar} \mathcal{W}_{\textrm{D}} \nonumber \\
  & \approx s\mathcal{P}\exp\Bigl\{\frac{i}{\hbar}\int_{0}^{T}
   d\tau\bigl[p\gamma_{0}-B_{\textrm{Ad}\mu}\dot{p}^{\mu}\bigr]\Bigr\}s^{-1}\,,
   \label{eq:adiabatic_def}
\end{align}
for adiabatic Berry phase depicted with $B_{\textrm{Ad}\mu}$. $s^{(-1)}$ are given in Eq.~\eqref{eq:sim_transform}.  Yet, in the worldline formalism in contrast to a quantum mechanical Berry phase we encounter two obstacles: 1) The propertime $T$ is a parameter, and moreover for full quantum observables is integrated over. Thus the large propertime criteria for adiabaticity in such instances is not met. 2) $p_\mu$ is also integrated over and moreover need not lie on shell; this distinction also impedes an adiabatic approximation.  Nevertheless,  for the construction of a classical kinetic theory--built from the classical equations of motion--such obstacles may be avoided since one may take the propertime integral which acts as a Lagrange multiplier enforcing the on-shell constraint. (We will also look at the quantum chiral anomaly case in Sec.~\ref{sec:vanishing}, in which the propertime criteria may surprisingly be met.) 
It is also of interest to speculate where appropriate the non-adiabatic case. This is for small gap and small (approaching the UV limit) propertimes such that $2pT\ll 2\pi \hbar$. We will,  however, mostly focus on the adiabatic Berry phase in this study so as to provide contrast from the worldline perspective to previous classical kinetic theory studies~\cite{Stephanov:2012ki,Chen:2013iga,Chen:2014cla,Stone:2013sga,Dwivedi:2013dea},  whom have all but exclusively employed the adiabatic theorem.

In order to identify the adiabatic part of the Berry phase for the Weyl representation gamma matrices let us show in matrix form the phase's explicit eigenspinor representation.  We have for $B_\mu \dot{p}^\mu$ the following
\begin{equation}
   \frac{\hbar}{2i}\gamma_{0}(\gamma_{0}-\gamma_{5})\begin{pmatrix}\bar{u}_{1}\dot{u}_{1} &   
   \bar{u}_{1}\dot{u}_{2} & \bar{u}_{1}\dot{v}_{1} & \bar{u}_{1}\dot{v}_{2}\\
   \bar{u}_{2}\dot{u}_{1} & \bar{u}_{2}\dot{u}_{2} & \bar{u}_{2}\dot{v}_{1} & \bar{u}_{2}\dot{v}_{1}\\
   \bar{v}_{1}\dot{u}_{1} & \bar{v}_{1}\dot{u}_{2} & \bar{v}_{1}\dot{v}_{1} & \bar{v}_{1}\dot{v}_{2}\\
   \bar{v}_{2}\dot{u}_{1} & \bar{v}_{2}\dot{u}_{2} & \bar{v}_{1}\dot{v}_{2} & \bar{v}_{2}\dot{v}_{2}
   \end{pmatrix}(\gamma_{0}-\gamma_{5})\,.
   \label{eq:phase_spelled_out}
\end{equation}
Then for large eigenvalue $\pm p$,  and hence large gap separating $u$ and $v$ leading to the adiabatic theorem,  we can determine the pieces that remain in the adiabatic limit are those with structure $\bar{u}_{i}\dot{u}_{j}$ and $\bar{v}_{i}\dot{v}_{j}$. One can compactly write the connection making use of the rotation components of the spin tensor, which are
\begin{equation}
   \Gamma_{\mu\nu}\coloneqq[\gamma_{\mu},\gamma_{\nu}]+\gamma_{0}[\gamma_{\mu},
   \gamma_{\nu}]\gamma_{0}\,,
   \label{eq:Gamma_Sigma}
\end{equation}
$\Gamma_{\mu\nu}$ is composed of two gamma matrices with non-equal spatial indices. Using the above one can find for the adiabatic Berry connection as
\begin{equation}
   B_{\textrm{Ad}}^{\mu}=i\frac{\hbar}{8}\frac{1}{p_{0}+p}\Gamma^{\mu\nu}\hat{p}_{\nu}\,,
   \label{eq:B_Ad}
\end{equation}
Note that, the connection here is block diagonal by virtue of the Weyl representation.

As anticipated for the adiabatic theorem with transformation to $p\gamma_{0}$ we have that $[B_{\textrm{Ad}}^{\mu},\gamma_{0}]=0$. Likewise one can find that $\{B^{\mu},\gamma_{0}\}=2B_{\textrm{Ad}}^{\mu}\gamma_{0}$.
The adiabatic Berry phase for Dirac fermions has been explored in~\cite{Mathur:1991ynd,Shankar:1994ta,Chang:2008zza}, where it was illustrated that the phase may likened to Thomas procession, with usage of Lorentz covariant spinors, and the spin-orbit interaction with usage of non-covariant ones. The adiabatic Dirac Berry phase in a quantum mechanical setting was further analyzed with motivation to a classical kinetic theory in~\cite{Chen:2013iga}.

Let us next address the Berry curvature.  First we confirm that the exact curvature disappears since it is a pure gauge, 
\begin{equation}
    S^{\mu\nu}_{\textrm{PG}}=\partial^{\mu}B^{\nu}_{\textrm{PG}}-\partial^{\nu}B^{\mu}_{\textrm{PG}}+i\hbar^{-1}[B^{\mu}_{\textrm{PG}},B^{\nu}_{\textrm{PG}}]=0\,,
\end{equation}
using Eq.~\eqref{eq:pure_gauge}. 
However the adiabatic curvature is finite. It is useful to introduce the following identity with the tensor given in Eq.~\eqref{eq:Gamma_Sigma}:
\begin{equation}
   [\Gamma_{\mu\nu},\Gamma_{\alpha\beta}]
   =8[w_{\nu[\alpha}\Gamma_{\mu\beta]}+w_{\mu[\beta}\Gamma_{\nu\alpha]}]\,,
\end{equation}
for $w_{\mu\nu}\coloneqq g_{\mu\nu}-g_{\mu0}g_{\nu0}$. Then, we can determine the adiabatic curvature as
\begin{align}
  S_{\textrm{Ad}}^{\mu\nu}&=\partial^{p\,\mu}B_{\textrm{Ad}}^{\nu}-\partial^{p\,\nu}B_{\textrm{Ad}}^{\mu}+i\hbar^{-1}[B_{\textrm{Ad}}^{\mu},B_{\textrm{Ad}}^{\nu}] \nonumber \\
  &=i\frac{\hbar}{8}\frac{(\hat{p}^{[\mu}+g^{[\mu0})p_{\alpha}
   \Gamma^{\alpha\nu]}-(p+p_{0})\Gamma^{\mu\nu}}{p^{2}(p_{0}+p)}
   \label{eq:S_Ad}\,,
\end{align}
We find the above connection and curvatures are radial~\cite{Shankar:1994ta}, i.e., $p_{\mu}B_{\textrm{Ad}}^{\mu}=p_{\mu}S_{\textrm{Ad}}^{\mu\nu}=0$.

In this section we have defined a Berry connection and curvature in the phase space worldline representation that serve as a natural extension from the quantum mechanical picture. An adiabaticity is also discussed, however we leave the study on the Berry phase for non-adiabatic processes for future work. Before we may construct a classical kinetic theory from the adiabatic Berry phase we must first employ a coherent state picture so that we can treat a well-defined scalar weighted worldline action. The coherent state picture is furthermore of value in its physical opaqueness.

\section{Worldline classical kinetic theory for Dirac fermions} \label{sec:QKT}

Having treated the Berry phase for Dirac fermions in the phase-space worldline setting,  inline for the Weyl case,  one can see that the adiabatic phase is non-Abelian. Therefore the worldline action is matrix weighted and a path ordering is present; (c.f., Eq.~\eqref{eq:Weyl_adiabatic}: the Weyl case possesses neither after the adiabatic theorem.) And determination of the Euler-Lagrange equations is challenging. To address this we make use of a coherent state, and fortunately a coherent state formalism exists for the worldline phase space representation under the initial work of Barut and Zanghi~\cite{PhysRevLett.52.2009}. And, the formalism has been applied to the path integral~\cite{Barut:1988ud,*Barut:1984vnk}. 

\subsection{Barut-Zanghi spinor coherent state} 
\label{sec:BZ_coherent}

With the BZ coherent state, the phase space conjugate variables, $(x_{\mu},p_{\mu})$, are enlarged to accommodate spin degrees of freedom with c-number spinors $(z,-i\bar{z})\in\mathbb{C}_{4}$, with $\bar{z}=z^\dagger \gamma_0$. BZ spinors exist in a first-quantized setting, however they possess many features in common with the usual second-quantized spinors in QFTs, we will illustrate below. 

First, to  get a better understand the formalism,  let us briefly review the case \textit{before} having applied the Hamiltonian transformation,  Eq.~\eqref{eq:Hamil_trans}, leading to Berry's phase. To do so let us begin by writing down the worldline action of Eq.~\eqref{eq:Eff_Action} with Eq.~\eqref{eq:S_A} with the BZ spinors as
\begin{equation}
    S_{\textrm{D}} = S_{\textrm{A}}+\int_{0}^{T}d\tau\bigl[\bar{z}\slashed{p}z+i\hbar\bar{z}\dot{z}\bigr]\,,
    \label{eq:S_D_before}
\end{equation}
where the $z,\bar{z}$ are the auxiliary c-number spinors acting on Dirac indices. One may find the equations of motion from the above action as 
\begin{equation}
   \dot{x}_{\mu}=\bar{z}\gamma_{\mu}z\,,\; \dot{p}_{\mu}=\frac{e}{c}F_{\mu\nu}\dot{x}^{\nu}\,,\;
   \dot{z}=\frac{i}{\hbar}\slashed{p}z\,,\; \dot{\bar{z}}=-\frac{i}{\hbar}\bar{z}\slashed{p}\,.
   \label{eq:eom_before}
\end{equation}
An advantage of the usage of BZ spinors comes from an intuitive interpretation of their bilinear form to the corresponding QFT observable. For example, consider the velocity with $\bar{z}\gamma_{\mu}z$; we can see this quantity may be likened to the vector current, however at classical level and without second quantized operators obeying the Clifford algebra. A symplectic system with appropriate Poisson brackets may also be constructed~\cite{PhysRevLett.52.2009}. Last, consider the case of no background electromagnetic field,  notice that an oscillating motion is present about the center-of-mass frame,  i.e.,  there are sinusoidal terms in $\dot{x}_{\mu}$ with argument $2p\tau$. This is the Zitterbewegung~\cite{schrodinger1930sitzungsber}, which predicts a rapid center of charge oscillation about the center-of-mass of the electron. More details for the above equations can be found in~\cite{PhysRevLett.52.2009,Barut:1988ud,*Barut:1984vnk}. 

\textit{After} having applied the transformation of Eq.~\eqref{eq:Hamil_trans} with $s$ given in Eq.~\eqref{eq:sim_transform} one can see 
\begin{equation}
    \bar{z}\slashed{p}z  \rightarrow\bar{z}s^{-1}\slashed{p}sz \,,\quad\bar{z}\dot{z} \rightarrow\bar{z}\dot{z}+\bar{z}s^{-1}\dot{s}z \,.
\end{equation}
One can then recognize the SO$(1,3)$ transformation as being a Lorentz transformation of the gamma matrices, characterizable for momentum dependent boost/rotation, $\omega_{\mu\nu}$, as 
$s(p)=\exp[(i/2)\omega_{\mu\nu}(p)\,\sigma^{\mu\nu}]$ and
$s^{-1}(p)=\exp[-(i/2)\omega_{\mu\nu}(p)\,\sigma^{\mu\nu}]$. 

To construct a path integral from a path ordered element such as Eq.~\eqref{eq:Wilson_Dirac} obeying the Dirac matrix representation of the Clifford group in 3+1 dimensions,  we require the resolution of the identity.  Consider for
$(z,-i\bar{z})\in\mathbb{C}_{4}$ 
the normalization 
\begin{equation}
    \mathcal{Z}_{0}=\int d\bar{z}dz\,e^{-\bar{z}z}=\pi^{4}\,,
\end{equation}
which one may supplement with generators such that
\begin{equation}
   \mathcal{Z}_{\eta}=\int dzd\bar{z}e^{-\bar{z}z+\bar{z}\eta+\bar{\eta}z}
   =\mathcal{Z}_{0}e^{\bar{\eta}\eta}\,.
\end{equation}
Then the resolution of the identity immediately follows as,
\begin{equation}
   \int d\Omega_{z}\,z_{a}\bar{z}_{b}\coloneqq\mathcal{Z}_{0}^{-1}\partial_{\bar{\eta}_{a}} 
   \partial_{\eta_{b}}\mathcal{Z}_{\eta}\Big|_{\eta=0}=\delta_{ab}\,,
   \label{eq:identity}
\end{equation}
where we have made explicit Dirac indices. Similar manipulations lead to 
\begin{align}
   \int d\Omega_{z}\,z_{a}\bar{z}_{b}z_{c}\bar{z}_{d} 
   &=\delta_{cd}\delta_{ab}+\delta_{ad}\delta_{bc}\,,\notag \\
   \int d\Omega_{z}\,z_{a}\bar{z}_{b}z_{c}\bar{z}_{d}z_{e}\bar{z}_{f} 
   &=\delta_{ab}\delta_{cd}\delta_{ef}+\delta_{ab}\delta_{cf}\delta_{ed} 
   +\delta_{ad}\delta_{bc}\delta_{ef} \notag \\
   +\delta_{ad}\delta_{cf}\delta_{be}&+\delta_{af}\delta_{bc}\delta_{ed}
   +\delta_{af}\delta_{cd}\delta_{be}\,,
   \label{eq:z_identities}
\end{align}
etc., corresponding to all permutations of spinor bilinears.

\subsection{Covariant and classical equations of motion with Berry phase}
\label{sec:CKT}

Equipped with a Dirac fermion Berry phase and also a spinor-like coherent state one can cast the path ordered factor into a path integral with Berry's phase. To achieve this let us break up the path ordered Dirac matrix weighted exponential as
\begin{equation}
   \mathrm{tr}\mathcal{W}_{D}=\lim_{N\rightarrow\infty}\mathrm{tr}\mathcal{P}\prod_{n=0}^{N-1}
   \Bigl[\mathbb{I}_4+\frac{i}{\hbar}\frac{\tau}{N}p_{\mu}(\tfrac{n\tau}{N})\gamma^{\mu}\Bigr]\,, 
   \label{eq:path_break_up}
\end{equation}
where into each infinitesimal element we introduce both the redefinition including the Berry phase through $ss^{-1}=\mathbb{I}_4$,  for similarity transform given by Eq.~\eqref{eq:sim_transform},  and also the resolution of unity of the coherent state,  Eq.~\eqref{eq:identity}.  Then taking the large $N$ limit the Dirac matrix exponential becomes
\begin{equation}
   \mathrm{tr}\mathcal{W}_{D}=\oint\mathcal{D}\Omega_{z}\exp\Bigl\{\frac{i}{\hbar}\int_{0}^{T}
   d\tau\bigl[\bar{z}p\gamma_{0}z-\bar{z}B_{\mu}z\dot{p}^{\mu}+i\hbar\bar{z}\dot{z}\bigr]\Bigr\}\,.   
   \label{eq:trWBerry}
\end{equation}
Here the $\oint\mathcal{D}\Omega_{z}$ describes the measure with the coherent state integral,  Eq.~\eqref{eq:identity},  with periodic boundary conditions, $z(0)=z(T)$,  by virtue of the Dirac trace. Note also that like before we assume here that $p(0)=p(T)$,  and hence $s(p(0))=s(p(T))$,  since one may equally well write the trace of the effective action in terms of momentum as opposed to the coordinates. 

Applying Eq.~\eqref{eq:trWBerry} to the effective action in  Eq.~\eqref{eq:Eff_Action}, we see that the worldline action now becomes
\begin{equation}
   S'_{\textrm{D}}=S_{\textrm{A}}+\int_{0}^{T}d\tau\bigr[\bar{z}p\gamma_{0}z
   -\bar{z}B_{\mu}z\dot{p}^{\mu}+i\hbar\bar{z}\dot{z}\bigl]\,,
\end{equation}
where $S_{\textrm{A}}$ is the part of the action given by Eq.~\eqref{eq:S_A}; c.f. see Eq.~\eqref{eq:S_D_before}. The equations of motion for enlarged phase space of $(x_{\mu},p_{\mu},z,\bar{z})$ are
\begin{align}
   \dot{x}_{\mu}&=\bar{\mathcal{D}}_{\mu}-\bar{S}_{\mu\nu}\dot{p}^{\nu}\,,\quad 
   \dot{p}_{\mu}=\frac{e}{c}F_{\mu\nu}\dot{x}^{\nu}\,, 
   \label{eq:x_eom}\\
   \dot{\bar{z}}&=-\frac{i}{\hbar}\bar{z}(p\gamma_{0}-B_{\mu}\dot{p}^{\mu})\,,\;
   \dot{z}=\frac{i}{\hbar}(p\gamma_{0}-B_{\mu}\dot{p}^{\mu})z\,,\label{eq:z_eom}
\end{align}
where we have denoted spin averaged quantites as 
$\bar{S}_{\mu\nu}\coloneqq\bar{z}S_{\mu\nu}z$,
$\bar{\mathcal{D}}_{\mu}\coloneqq\bar{z}\mathcal{D}_{\mu}z$, 
etc.  Here the diagonal element with commutation with the Berry phase reads 
\begin{equation}
   \mathcal{D}_{\mu}\coloneqq\gamma_{0}\hat{p}_{\mu}-i\hbar^{-1}p[\gamma_{0},B_{\mu}]\,.
\end{equation}
Note that the commutation term vanishes for the adiabatic Berry connection.
The above equations are the 3+1 dimensional Lorentz covariant extension in phase space of the 3 dimensional ones found in Eq.~\eqref{eq:weyl_current_berry}, or in~\cite{Stephanov:2012ki, Chen:2013iga,Chen:2014cla}.

We must invert 
\begin{equation}
   \mathcal{G}_{\mu\nu}\coloneqq g_{\mu\nu}+\frac{e}{c}\bar{S}_{\mu\sigma}F_{\;\nu}^{\sigma}\,,   
   \label{eq:A}
\end{equation}
and it is convenient to do the inversion with the help of Cayley Hamilton's theorem.  A few Lorentz invariants will come in handy: 
\begin{equation}
   I_{\tilde{F}F}=-\frac{1}{4}\widetilde{F}_{\mu\nu}F^{\mu\nu},\;
   I_{\tilde{\bar{S}}\bar{S}}=-\frac{1}{4}\widetilde{\bar{S}}_{\mu\nu}\bar{S}^{\mu\nu},\;
   I_{\bar{S}F}=\frac{1}{2}\bar{S}_{\mu\nu}F^{\mu\nu}.
   \label{eq:Lorentz_invariants}
\end{equation}
where $\widetilde{F}_{\mu\nu}=(1/2)\epsilon_{\mu\nu\alpha\beta}F^{\alpha\beta}$.  Let us also introduce the following identities, 
\begin{align}
   &F^2-\widetilde{F}^2=-I_{\tilde{F}F} \mathrm{I}_4\,,\;
   \bar{S}^2-\widetilde{\bar{S}}^2=-I_{\tilde{\bar{S}}\bar{S}} \mathrm{I}_4 \,,\notag\\
   &\widetilde{F} F =I_{\tilde{F}F} \mathrm{I}_4 \,,\;
   \widetilde{\bar{S}} \bar{S} =I_{\tilde{\bar{S}}\bar{S}}\mathrm{I}_4 \,,\;
   \bar{S} F -\widetilde{F} \widetilde{\bar{S}} =-I_{\bar{S}F} \mathrm{I}_4\,,
\end{align}
where we use an implicit Lorentz index matrix notation, e.g., $F^\mu_{\;\nu}\eqqcolon F$.
Then we can find that 
\begin{equation}
    (\bar{S} F)^2 +I_{\bar{S}F}\bar{S} F -I_{\tilde{F}F}I_{\tilde{\bar{S}}\bar{S}} \mathrm{I}_4=0\;,
\end{equation}
which is Cayley Hamilton's theorem stemming from 
$\det[\mathcal{G}-\lambda\delta]$
for generic eigenvalue $\lambda$. Then using the above formula,  one can verify the inverse of Eq.~\eqref{eq:A} is
\begin{equation}
   [\mathcal{G}^{-1}]_{\mu\nu}=\frac{1}{\sqrt{\det\mathcal{G}}}\Bigl[g_{\mu\nu}
   +\frac{e}{c}\widetilde{F}_{\mu\sigma}\widetilde{\bar{S}}_{\;\nu}^{\sigma}\Bigr]\,,
\end{equation}
where $\sqrt{\det\mathcal{G}}=1-(e/c)I_{\bar{S}F}-(e/c)^{2}I_{\tilde{F}F}I_{\tilde{\bar{S}}\bar{S}}$.
Then we can then find for the equations of motion,  Eqs.~\eqref{eq:x_eom}-\eqref{eq:z_eom},
\begin{align}
   \sqrt{\det\mathcal{G}}\dot{x}_{\mu}&=\Bigl[g_{\mu\nu}
   +\frac{e}{c}\widetilde{F}_{\mu\sigma}\widetilde{\bar{S}}_{\;\nu}^{\sigma}\Bigr]\bar{\mathcal{D}}^{\nu}\,,
   \label{eq:worldline_cke_eom1}\\
   \sqrt{\det\mathcal{G}}\dot{p}_{\mu}&=\Bigl[\frac{e}{c}F_{\mu\nu}
   +\Bigl(\frac{e}{c}\Bigr)^{2}I_{\tilde{F}F}\widetilde{\bar{S}}_{\mu\nu}\Bigr]\bar{\mathcal{D}}^{\nu}\,.
   \label{eq:worldline_cke_eom2}
\end{align}
These are the covariant worldline equivalent of Eq.~\eqref{eq:weyl_current_berry},
(or those in a quantum mechanical setting as shown in~\cite{Stephanov:2012ki, Chen:2013iga, Chen:2014cla}).

Let us first remark on the exact pure gauge case, Eq.~\eqref{eq:pure_gauge}.  It can be clearly seen that only a trivial correction to the equations of motion can be found, i.e., $\dot{x}_\mu = \bar{\mathcal{D}}_\mu$ and $\dot{p}_\mu=(e/c)F_{\mu\nu}\bar{\mathcal{D}}^\nu$, since the Berry curvature vanishes in the pure gauge case.  There will be no terms that contain the anomaly factor, $I_{\tilde{F}F}$,  and hence one cannot construct a classical kinetic theory.  Therefore, let us examine just the adiabatic case for a classical kinetic theory.

Similar to the Weyl case,  one may find for the modified and canonically conserved~\cite{PhysRevLett.96.099701} phase space measure as
\begin{equation}
   d\mu_{\textrm{D}}=\sqrt{\det\mathcal{G}}\frac{d^{4}pd^{4}x
   d\Omega_{z}}{(2\pi)^{4}}\,.
   \label{eq:can_measure}
\end{equation}
The phase space measure has been extended to include BZ spinors. To construct a classical kinetic theory let us define a phase space distribution function, $f$, that satisfies a collisionless Boltzmann equation,
\begin{equation}
   \frac{d}{d\tau}f=\Bigl[\frac{\partial}{\partial\tau}+\dot{x}^{\mu}\partial_{\mu}
   +\dot{p}^{\mu}\partial_{\mu}^{p}+\dot{z}_{a}\partial_{a}^{z}
   +\dot{\bar{z}}_{a}\partial_{a}^{\bar{z}}\Bigr]f=0\,.
   \label{eq:Boltzmann}
\end{equation}
Let us furthermore assume the most general distribution function,  which has the form of a Wigner function~\cite{Gao:2017gfq,Hattori:2019ahi,Hidaka:2022dmn} up to bilinear in BZ spinors,  and may be written as
\begin{align}
   f&(x,p,z,\bar{z})= f_0(x,p)\notag \\
   &+\frac{1}{4}\bar{z}\bigl\{\mathcal{S}
   +\gamma_{5}\mathcal{P}+\gamma_{\mu}\mathcal{V}^{\mu}
   +\gamma_{\mu}\gamma_{5}\mathcal{A}^{\mu}
   +\sigma_{\mu\nu}\mathcal{T}^{\mu\nu}\bigr\}z\,.
   \label{eq:dist}
\end{align}
The key distinction between the above and the distribution function for the Weyl case is that we no longer have pure left or right handed distributions. Therefore averaging $\dot{x}$ with Eq.~\eqref{eq:dist} over momentum and spin no longer represents a vector current density. However, by the physical opaqueness of the BZ formalism we can interpret the classical quantities. This is simplest to do studying the case \textit{before} having applied the Hamiltonian transformation leading to the Berry phase with classical equations of motion given in Eq.~\eqref{eq:eom_before}--here the canonically conserved phase space measure is Eq.~\eqref{eq:can_measure} but with $\sqrt{\det\mathcal{G}}\rightarrow 1$. Then it can be readily inferred after averaging over the BZ spin variables, using Eq.~\eqref{eq:z_identities}, that each constituent in Eq.~\eqref{eq:dist} may be likened to its associated quantum expectation value. Namely we have that the vector and axial vector currents are respectively given by
\begin{align}
    j_{\mathcal{V}\mu}&=\int\frac{d^{4}pd\Omega_{z}}{4(2\pi)^{4}}\sqrt{\det\mathcal{G}}\,\dot{x}_{\mu}\,\bar{z}\gamma_{\nu}\mathcal{V}^{\nu}z\,,\\
    j_{\mathcal{A}\mu}&=\int\frac{d^{4}pd\Omega_{z}}{4(2\pi)^{4}}\sqrt{\det\mathcal{G}}\,\dot{x}_{\mu}\,\bar{z}\gamma_{\nu}\gamma_5\mathcal{A}^{\nu}z\,.
\end{align}
Classical currents associated with $f_0$, $\mathcal{S}$, $\mathcal{P}$, and $\mathcal{T}^{\mu\nu}$ vanish after taking the BZ spinor integral for either the \textit{before} (with velocity given in Eq.~\eqref{eq:eom_before}) or \textit{after} (Eqs.~\eqref{eq:worldline_cke_eom1}-\eqref{eq:worldline_cke_eom2}) Berry phase cases due to a trace over an odd number of gamma matrices. What is interesting but anticipated is that $j_{\mathcal{A}\mu}$ vanishes \textit{before} having taken the transformation leading to the Berry phase; let us now examine \textit{after} the transformation.

To arrive at a classical phenomenological equivalent of the anomaly let us make use of the fact that
\begin{align}
   &\partial_{\mu}(\sqrt{\det\mathcal{G}}\dot{x}^{\mu})=0\,,\\
   &\partial_{p}^\mu(\sqrt{\det\mathcal{G}}\dot{p}_\mu)
   =\left(\frac{e}{c}\right)^{2}I_{\tilde{F}F}\partial_{p}^{\mu}
   (\widetilde{\bar{S}}_{\mu\nu}\bar{\mathcal{D}}^{\nu})\,,\\
   &\partial_{z\,a}(\sqrt{\det\mathcal{G}}\dot{z}_{a})
   +\partial_{\bar{z}\,a}(\sqrt{\det\mathcal{G}}\dot{\bar{z}}_{a}) \notag\\
   &\;=\frac{i}{2\hbar}\frac{e}{c}\left(\frac{e}{c}I_{\tilde{F}F}\widetilde{\bar{S}}_{\mu\nu} 
   -F_{\mu\nu}\right)\bar{z}[S^{\mu\nu},p\gamma_{0}-B_{\mu}\dot{p}^{\mu}]z\,,
\end{align}
Also, our treatment so far has been generic. Let us now go ahead and assume an adiabatic connection, Eq.~\eqref{eq:B_Ad},  and curvature,  Eq.~\eqref{eq:S_Ad}; also $\mathcal{D}_{\mu}\rightarrow\gamma_{0}\hat{p}_{\mu}$ in the adiabatic case.  Then, using Eqs.~\eqref{eq:z_identities}, we may go ahead and integrate out the spin degrees of freedom to find as anticipated that the vector current is conserved,
\begin{equation}
   \partial^\mu j_{\mathcal{V}\mu}=0\,.
\end{equation}
However the divergence of the classical axial vector current can be found as
\begin{align}
    \partial^\mu j_{\mathcal{A}\mu}&= \Bigl(\frac{e}{c}\Bigr)^{2}I_{\tilde{F}F}\int\frac{d^4p}{(2\pi)^4}\frac{1}{2}\textrm{tr}
   \gamma_{\rho}\gamma_{5}\mathcal{A}^{\rho}\partial^{p}_{\mu}\widetilde{S}_{\textrm{Ad}}^{\mu\nu}
   \gamma_{0}\hat{p}_{\nu}\notag\\
   &=\Bigl(\frac{e}{c}\Bigr)^{2}I_{\tilde{F}F}\int\frac{d^4p}{(2\pi)^4}\frac{-2\hbar}{p^{3}(p_{0}+p)}\boldsymbol{\mathcal{A}}\cdot\boldsymbol{p}\,.
   \label{eq:ckt_nonconservation}
\end{align}
We find the non-conservation of classical axial current is proportional to the factor of the anomaly, $I_{\tilde{F}F}$, Eq.~\eqref{eq:Lorentz_invariants}, as expected, indicating an anomalous contribution at a classical level. And an initial axial vector distribution, $\mathcal{A}^\mu$, must be present for nonvanishing Eq.~\eqref{eq:ckt_nonconservation}; in this way a net chirality may be introduced. In a similar way, for the Weyl case; one strictly had a distribution for a chiral particle with enslaved helicity owing to the adiabaticity,  and in order for the classical axial anomaly to appear the degenerate gas of Weyl fermions had to have an imbalance in chemical potentials for the chiral particles. Finally, forms of $\mathcal{A}^\mu$ can be reasoned such that the right-hand side of Eq.~\eqref{eq:ckt_nonconservation} be equivalent to that given by the axial anomaly, namely with value $\hbar e^2/(2c^2\pi^2) I_{\tilde{F}F}$. However, the axial vector distribution is largely unconstrained, with exception to the Boltzmann equation in Eq.~\eqref{eq:Boltzmann}, therefore we leave the expression in  Eq.~\eqref{eq:ckt_nonconservation} as its final form.

Classical constructions of the anomaly in a kinetic theory have been explored for Weyl fermions~\cite{Stone:2013sga,Dwivedi:2013dea}, and also for massive and massless Dirac fermions in~\cite{Chen:2013iga}. Notably in~\cite{Chen:2013iga} through a quantum mechanical non-Abelian Berry phase it was found the anomaly appeared only after application of the massless limit. At the quantum level, particularly with utilization of the Wigner function formalism, kinetic theories that encapsule the axial anomaly have been well-studied; see Sec.~\ref{sec:intro} for pertinent literature. Let us point out, that our setup in contrast with others' which make use of a Wigner function formalism possess differences.  Most notably we must insert a similarity transform to arrive at the Berry phase and hence classical description of a classical kinetic theory.  If such a similarity transform was introduced into, e.g., a Schwinger-Keldysh partition function, the initial density matrix would be transformed as well. Therefore, one must inherently treat the classical description of the axial current nonconservation in Eq.~\eqref{eq:ckt_nonconservation} at a phenomenological level.

We will expose in the next section a striking example of why Eq.~\eqref{eq:ckt_nonconservation} cannot fully represent the axial anomaly at the quantum level: We will apply the adiabatic approximation to the quantum anomaly to show the anomaly always vanishes.

\section{Vanishing index under adiabaticity}
\label{sec:vanishing}

We saw above that through an incompressible phase space by virtue of the Berry phase,  what would be a quantum level crossing was imparted into an otherwise classical construction--just as was originally demonstrated in~\cite{Stephanov:2012ki,Chen:2013iga,Chen:2014cla},  and for the Weyl case in Sec.~\ref{sec:Weyl}.  Also like the Weyl case where a chiral chemical potential was used,  for the Dirac case in order to see a non-conservation of the current a distribution function with axial vector coupling was required.  However,  the quantum anomaly exists independent of an initial distribution, and no phenomenological addition is needed, therefore it is prudent we explore the fully quantum case with appeal to a Berry phase.  We will demonstrate that the axial anomaly under the adiabatic approximation as conceived in Eq.~\eqref{eq:adiabatic_def} for Dirac fermions in a phase space worldline setting vanishes.  

Let us treat the index theorem for just such a quantum description~\cite{tHooft:1976snw,*Atiyah:1978ri}.  It has the virtue of describing rigorously the sum over the zero modes of the Dirac operator in a gauge field. We may equate the index to the number of chiral modes and hence the quantum anomaly description of a non-conservation of axial current for massless fermions. Our strategy will be to apply the Berry transformation and adiabatic theorem; the assumption is if an adiabatic Berry phase as led to in Eq.~\eqref{eq:adiabatic_def} shares a topological origin to the index theorem,  then such a step would be unhindered, otherwise their connection would be null.  Let us emphasize we treat only the adiabatic case,  taking an exact pure gauge transformation Eq~\eqref{eq:pure_gauge} would not hinder the index theorem.

We take as our definition of the index in Minkowski spacetime:
\begin{equation}
   I_n\coloneqq\lim_{M\rightarrow\infty}\textrm{Tr}\,\gamma_{5}\frac{-M}{i\hbar\slashed{D}-M}\,,
   \label{eq:index_theorem}
\end{equation}
which agrees with the definition for the pseudoscalar condensate for large mass.  To arrive at a more conventional definition~\cite{Vandoren:2008xg} note that also 
$I_n=\lim_{M\rightarrow\infty}\textrm{Tr}\,\gamma_{5}[M^{2}/(\hbar^2\slashed{D}^2+M^{2})]$.
One may cast the above into a phase space worldline path integral:
\begin{equation}
   I_n=\lim_{M\rightarrow\infty}\mathrm{tr}\int d^{4}p\,M\gamma_{5}\,G(A,p,p)\,,
\end{equation}
with Green's function given by Eq.~\eqref{eq:Green} with $m\rightarrow M$ written in the momentum representation.  

Now let us turn our attention to the evaluation of the traced path ordered element; this is 
$\mathrm{tr}\gamma_{5}\mathcal{W}_{\textrm{D}}$.
One may break up path ordered element as shown in Eq.~\eqref{eq:path_break_up},  inserting in a complete sets of states,  i.e., $ss^{-1}=\textrm{I}_4$ with $s$ given by Eq.~\eqref{eq:sim_transform} to arrive at 
\begin{equation}
   \mathrm{tr}\gamma_{5}\mathcal{W}_{\textrm{D}}
   =\mathrm{tr}\gamma_{5}\mathcal{P}\exp\Bigl\{\frac{i}{\hbar}\int_{0}^{T}d\tau\,
   [\gamma_{0}p-B_{\mu}\dot{p}^{\mu}]\Bigr\}\,,
   \label{eq:index_Berry}
\end{equation}
where we note that 
$[s,\gamma_{5}]=[s^{-1},\gamma_{5}]=0$ 
and that $s(p(0))=s(p(T))$ due to the periodicity criteria. 

That we may apply the adiabatic theorem and drop level crossing terms amounts to the propertime, $T$, and $p$ be finite and large; see Eq.~\eqref{eq:adiabatic_def} and related arguments in Sec.~\ref{sec:DiracBerry}.  Specifically one has $pT\gg \pi \hbar$.  As an initial observation,  let us absorb $M$ into the Schwinger proper time,  $T$,  such that 
$T\rightarrow M^{-1}T$. 
This step would already be at odds with the adiabaticity criteria sending the integration limits to $[0,M^{-1}T)$,  which would negate the criteria for the adiabatic theorem.  Nevertheless, one may show that 
$\mathrm{tr}\gamma_{5}\mathcal{W}_{\textrm{D}}$
is in fact independent of $T$ since
\begin{align}
   \frac{\hbar}{i}\frac{d}{dT}\textrm{tr}\gamma_{5}\mathcal{W}_{\textrm{D}}
  &=\textrm{tr}\gamma_{5}\slashed{p}(T)\mathcal{W}_{\textrm{D}}
   =\textrm{tr}\gamma_{5}\mathcal{W}_{\textrm{D}}\slashed{p}(0)\notag \\
   &=-\textrm{tr}\gamma_{5}\slashed{p}(T)\mathcal{W}_{\textrm{D}}=0\,,
  \label{eq:IT_indep}
\end{align}
where we have made use of the definition of the path ordering in $T$,  the Dirac trace,  and the momentum periodicity criteria: $\slashed{p}(T)$ may be put into the path ordering because it is already in its path ordered place.  However since $p(0)=p(T)$, it may equally well go to the front of the path ordering.  The argument is analogous to the fact that $\partial I_n/\partial M^{2}=0$ for finite $M$ in Eq.~\eqref{eq:index_theorem}; see~\cite{Vandoren:2008xg} and references therein for details.  Therefore,  Eq.~\eqref{eq:index_Berry} too must be independent of $T$.  Even so we surprisingly find that with the adiabatic approximation the index vanishes.  Namely
\begin{equation}
   \mathrm{tr}\gamma_{5}\mathcal{P}\exp\Bigl\{\frac{i}{\hbar}\int_{0}^{T}d\tau\,
   [\gamma_{0}p-B_{\textrm{Ad}\mu}\dot{p}^{\mu}]\Bigr\}=0\,,
\end{equation}
and hence
\begin{equation}
   \lim_{\textrm{Ad}}I_n = 0\,,
   \label{eq:vanishing}
\end{equation}
where the limit acts on the path ordered element as shown in Eq.~\eqref{eq:adiabatic_def}. This step can be clearly seen by expanding the path ordering in the above similar as was defined for Eq.~\eqref{eq:path_break_up}.  Since $B_{\textrm{Ad}}^{\mu}$ will always be a product of two spatial gamma matrices, i.e. , $\gamma_{i}\gamma_{j}$ for $i,j=1-3$,  one will find an infinite product of \textit{even} spatial gamma matrices,  and a product of $\gamma_0$ matrices.  Hence, the trace with $\gamma_{5}$ must always vanish.  Therefore,  the adiabatic approximation for the index theorem cannot be applied.

This statement holds quite generally.  Even if we selected, instead of $s$ given in Eq.~\eqref{eq:sim_transform}, $s$ with spinors such that $s^{-1}\slashed{p}s = \gamma_5 p$, one would find the same disappearance.  Such a transform could be written $(1/\sqrt{2}) [u_1,u_2,v_1,v_2]$.  We will further look at another example of a similarity transform in the following section not built from covariant spinors, but which still follows the same adiabaticity criteria; using it too one can see that Eq.~\eqref{eq:vanishing} still holds.

While the propertime criteria for the adiabatic theorem can be met according to Eq.~\eqref{eq:IT_indep},  finite (large) $p$ cannot since all values are integrated over.  Then when the adiabatic theorem is applied, a level crossing at $p=0$ is inhibited. However such a level crossing makes possible the anomaly in topological gauge fields,  and therefore by applying the adiabatic theorem we have inhibited the anomaly and hence why it is thought Eq.~\eqref{eq:vanishing} vanishes.  We can furthermore stress this inhibition noting that the index, according to Eq.~\eqref{eq:IT_indep}, is independent of the mass, $M$ since $\mathrm{tr}\gamma_{5}\mathcal{W}_{\textrm{D}}$ is independent of $T$,  which would have enforced a coupling to the mass.  Since there is no mass coupling the adiabatic theorem, relying on both a large mass and on-shell criteria,  is incompatible with the index. Let us also remark that a similar phenomenon of a vanishing anomaly has been found in~\cite{Copinger:2018ftr}, whereby the divergence of the axial current, through the axial Ward identity (comprised of the index and pseudoscalar condensate components), vanishes.  Last,  that the adiabatic theorem cannot lead to the quantum anomaly was first examined in~\cite{Deguchi:2005pc,*Fujikawa:2005cn}. However, our approach here is novel in that we may directly contrast with the kinetic theory through the adiabatic Berry phase. 

A curious corollary follows from the above arguments: instead of the adiabatic approximation, one uses the off-diagonal level crossing components of the Berry phase, i.e., those with $\bar{v}\dot{u}$ or $\bar{u}\dot{v}$ in Eq.~\eqref{eq:phase_spelled_out}; then $I_n$ need not disappear from taking the Dirac trace.  Exploring this more fully is, however, left for future studies.

While we have seen that the adiabatic Berry phase makes the index, or quantum anomaly, vanish, the same cannot be said for an exact Berry phase,  or a transformation given by Eq.~\eqref{eq:Hamil_trans}. With the aid of the Fujikawa method applied to the BZ coherent state one can observe several commonalities between both chiral rotations and transformations leading to an exact Berry phase.

\section{Fujikawa method on the worldline}
\label{sec:fujikawa}

To begin our discussion let us first show how an axial rotation in the BZ spinor coherent state can reproduce the axial-Ward identity~\cite{Adler:1969gk,*Bell:1969ts} via the Fujikawa method~\cite{Fujikawa:1979ay,*Fujikawa:1980eg}.  
This derivation serves two purposes: We may shortly contrast the axial rotation with a similarity transform leading to the Berry phase.  And we may also confirm the validity of the BZ spinor coherent state,  especially confirming how the formalism relates to the subtleties of an axial rotation to the fermionic determinant.
To begin let us take the effective action, Eq.~\eqref{eq:Eff_Action},  however written with the aid of the coherent state:
\begin{align}
   \Gamma[A]&=i\hbar\int_{0}^{\infty}\frac{dT}{T}\oint\mathcal{D}x\int\frac{\mathcal{D}p}{2\pi\hbar}
   \,e^{\frac{i}{\hbar}S_{\mathcal{A}}}\mathrm{tr}\mathcal{W}_{\textrm{D}}\,, \\
   \mathrm{tr}\mathcal{W}_{\textrm{D}}&= \oint\mathcal{D}\Omega_{z}\,  \exp\Bigl\{\frac{i}{\hbar}\int_{0}^{T}d\tau
   \bigl[\bar{z}\slashed{p}z+i\hbar\bar{z}\dot{z}\bigr]\Bigr\}\,,
   \label{eq:eff_action_z}
\end{align}
where $S_{\textrm{A}}$ can be found from Eq.~\eqref{eq:S_A}.  Then we perform the axial rotation
\begin{equation}
   z\rightarrow\exp[i\theta(\tau)\gamma_{5}]z\,.
   \label{eq:axial_rot}
\end{equation}
We acquire an anomalous phase from the transformation,  which in principal requires regularization for its evaluation. The BZ coherent state factor becomes
\begin{align}
   &\mathrm{tr}\mathcal{W}_{\textrm{D}}=\det\bigl[e^{2i\theta\gamma_{5}}\bigr]  \oint\mathcal{D}   
   \Omega_{z}\notag \\
    &\times \exp\Bigl\{\frac{i}{\hbar}\int_{0}^{T}d\tau\bigl[\bar{z}\slashed{p}z-\hbar\dot{\theta}\bar{z}   
    e^{2i\theta\gamma_{5}}\gamma_{5}z+i\hbar\bar{z}e^{2i\theta\gamma_{5}}\dot{z}\bigr]\Bigr\}\,.
    \label{eq:anomalous_phase}
\end{align}
Note that we have bosonic degrees of freedom for the coherent state,  and hence the sign of the determinant.

Rather than evaluating the anomalous phase directly,  we absorb it into functional determinant of the coherent state.  For the c-number spinor variable one finds
\begin{equation}
   \mathrm{tr}\mathcal{W}_{\textrm{D}}=
   \det\Bigl[e^{-2i\theta\gamma_{5}}\slashed{p}-\hbar\dot{\theta}\gamma_{5}
   +i\hbar\frac{d}{d\tau}\Bigr]^{-1}\,.
   \label{eq:eff_action_z2}
\end{equation}
One can also find the same functional determinant after integrating out the spinors from Eq.~\eqref{eq:eff_action_z},  then inserting $\exp(i\theta\gamma_{5})\exp(-i\theta\gamma_{5})$ into the argument of the determinant. 

Let us pause to notice the axial rotation can be written into the form of a Hamiltonian transformation via Eq.~\eqref{eq:Hamil_trans} as
\begin{equation}
   \slashed{p}\rightarrow e^{-i\theta\gamma_{5}}\slashed{p}e^{i\theta\gamma_{5}}
   +i\hbar e^{-i\theta\gamma_{5}}\frac{d}{d\tau}e^{i\theta\gamma_{5}}\,.
\end{equation}
Therefore,  one can classify an axial gauge transformation as an exact Berry or geometric phase according to Eq.~\eqref{eq:Hamil_trans}.  However, unlike the Berry phase where commonly evoked,  there is obviously no eigendecomposition in the axial transform.

To arrive at the axial-Ward identity let us reverse the steps we have taken with Eq.~\eqref{eq:eff_action_z2} to arrive at a path ordered expression,
\begin{equation}
   \mathrm{tr}\mathcal{W}_{\textrm{D}}= \mathrm{tr} \mathcal{P}
   \exp\Bigl\{\frac{i}{\hbar}\int_{0}^{T}d\tau
   \bigl[e^{-2i\theta\gamma_{5}}\slashed{p}-\hbar\dot{\theta}\gamma_{5}\bigr]\Bigr\}\,.
\end{equation}
Finally,  let us restrict the form of the rotation angle so that $\theta(\tau)=\theta(x(\tau))$, then the effective action may written as
\begin{equation}
   \Gamma[A]=-2\hbar\mathrm{Tr}\theta\gamma_{5}
   -i\hbar\textrm{Tr}\ln\Bigl[\slashed{p}-\frac{e}{c}\slashed{A}
   -\hbar\slashed{\partial}\theta\gamma_{5}-e^{2i\theta\gamma_{5}}mc\Bigr]\,,
\end{equation}
which is precisely the form of the QED partition function after applying the axial rotation. The axial-Ward identity can be found by looking at small $\theta$, and also by evaluating the functional trace of $\gamma_{5}$~\cite{Fujikawa:1979ay,*Fujikawa:1980eg}.

Having demonstrated how the Fujikawa method may be applied to the BZ spinor coherent state on the worldline,  let us look at a special eigendecomposition that unlike that used in Sec.~\ref{sec:DiracBerry} is composed of non-covariant eigenvectors. A non-covariant treatment for the Berry phase has also been explored in~\cite{Stone:2014fja}, which examined the phase's relationship to a spin-orbit coupling. The treatment has been explored in~\cite{Shankar:1994ta} where the Berry curvature was argued to possess topological characteristics of the Yang-Mills meron~\cite{DEALFARO1976163}.

Consider a similarity transform, $\tilde{s}$,  such that $\tilde{s}^{-1}\slashed{p}\tilde{s}=\gamma_{5}p$, but with
$\tilde{s}^{-1}\neq\gamma_{0}\tilde{s}^{\dagger}\gamma_{0}$, and hence cannot be cast into an SO$(1,3)$ transformation with Dirac matrix representation. The similarity transform reads
\begin{align}
   \tilde{s}&=\frac{1}{\sqrt{2}}(\mathbb{I}_4-\gamma_{5}\slashed{\hat{p}})
   =\exp \Bigl[-\frac{8n+1}{4}\pi\gamma_{5}\slashed{\hat{p}} \Bigr]\,,\label{eq:cliff_trans} \\
   \tilde{s}^{-1}&=\frac{1}{\sqrt{2}}(\mathbb{I}_4+\gamma_{5}\slashed{\hat{p}})
   =\exp \Bigl[ \frac{8n+1}{4}\pi\gamma_{5}\slashed{\hat{p}}\Bigr]\,,
\end{align}
for $n\in\mathcal{Z}$. 
The similarity transform resembles the axial transform, Eq.~\eqref{eq:axial_rot}, however with the additional placement of the linear term $\slashed{\hat{p}}$. The Berry phase here reads 
\begin{equation}
   \tilde{B}_{\textrm{PG}\mu}=\frac{1}{2p^2}[\sigma_{\mu\nu}p^{\nu}+i\gamma_5 (p\gamma_{\mu}-\hat{p}_\mu\slashed{p})]\,,
\end{equation}
which is a pure gauge transformation and the adiabatic approximation has not been used. A spin-tensor structure of the adiabatic phase, the $\sigma_{\mu\nu}p^{\nu}/(2p^{2})$ part of the connection, here is readily apparent. Note, we have that the phase here too is radial: $\tilde{B}_{\textrm{PG}\mu}p^\mu=0$.

Let us remark that application of Eq.~\eqref{eq:cliff_trans} instead of Eq.~\eqref{eq:sim_transform} to the path ordered element for the quantum axial anomaly presented in Sec.~\ref{sec:vanishing} would also lead to a vanishing index theorem. Therefore the incompatibility of the worldline adiabaticity and the index theorem is quite robust.

Similar to the above case with an axial rotation let us redefine the coherent state spinors with the transform given in Eq.~\eqref{eq:cliff_trans},
\begin{equation}
   z\rightarrow\tilde{s}z\,.
\end{equation}
Then in analogy to Eq.~\eqref{eq:anomalous_phase} we find for the BZ coherent state factor
\begin{align}
   &\mathrm{tr}\mathcal{W}_{\textrm{D}}=\det\bigl[e^{-\frac{(8n+1)}{2}\pi\gamma_{5}
   \hat{\slashed{p}}}\bigr]\oint\mathcal{D}\Omega_{z}\notag\\
    &\times \exp\Bigl\{\frac{i}{\hbar}\int_{0}^{T}d\tau\bigl[\bar{z}\slashed{p}z+i\hbar\bar{z}\tilde{s}
   \dot{\tilde{s}}z+i\hbar\bar{z}e^{-\frac{(8n+1)}{2}\gamma_{5}\hat{\slashed{p}}}\dot{z}\bigr]\Bigr\}\,,
   \label{eq:sim_anomalous}
\end{align}
where we have used the fact that 
$\tilde{s}^{\dagger}\gamma_{0}=\gamma_{0}\tilde{s}$, and that 
$\tilde{s}\slashed{p}\tilde{s}=\slashed{p}$. 
We find in a similar way to the axial rotation of Eq.~\eqref{eq:axial_rot}, here too lies a functional determinant stemming from the Jacobian of the transformation, however here with a functional trace of $\gamma_5 \slashed{\hat{p}}$. Also analogous to Eq.~\eqref{eq:anomalous_phase}, we see the similarity transform does not change the $\slashed{p}$ term. 

Following analogous steps as used for the axial rotation one can also find that
\begin{equation}
   \mathrm{tr}\mathcal{W}_{\textrm{D}}=\oint\mathcal{D}\Omega_{z}\exp\Bigl\{\frac{i}{\hbar}\int_{0}^{T}\hspace{-0.5em}d\tau[\bar{z}p\gamma_{5}z
   -\bar{z}\tilde{B}_{\textrm{PG}}^\mu z\dot{p}_{\mu}+i\hbar\bar{z}\dot{z}]\Bigr\}\,,
\end{equation}
which is the expected Berry phase transformation. An interesting merit of using the non-covariant eigenvector transformation can be seen in that now the worldline Lagrangian is manifestly Lorentz invariant and no side-jump~\cite{Chen:2014cla,Hidaka:2016yjf} term is present after boosting. 

Unlike the axial rotation case, here the passage to an equivalent effective action is not a straightforward procedure. Because of a the $\dot{p}^\mu$ term present in the Lagrangian a non-commutative structure persists, marring a simple form for the effective action. Therefore we cannot rigorously define the determinant given in Eq.~\eqref{eq:sim_anomalous} in an operator form, and hence cannot evaluate it at this time.

\section{Conclusions}

The Berry phase has been examined in the worldline formalism in a phase space representation with both aim to classical and quantum phenomena related to the axial anomaly. For the former classical kinetic theories were constructed for both Weyl and Dirac fermions. For the case of Weyl fermions the chiral kinetic theory in the phase space worldline representation resembled closely the same theory which could be found from a reduced dimension quantum mechanical construction. However, for Dirac fermions a non-Abelian Berry phase was present, which demanded application of a coherent state--we adopted spinors introduced by 
BZ--to the path ordered element~\eqref{eq:Wilson_Dirac} so that classical equation of motions could be found. With application of the adiabatic theorem on the phase space worldline, and introduction of a Wigner-like distribution function, a classical and covariant kinetic theory was formulated, whose non-conservation of axial current could be seen with axial vector contributions in the distribution function.

However, using the same adiabaticity that was applied to construct the Dirac classical kinetic theory, yet applied to the quantum Dirac operator index, it was found that the index vanished, suggesting that one may not apply the adiabatic theorem in determination of the axial anomaly. Therefore, we argue the anomalous effects found in the Dirac classical kinetic theory should be of phenomenological origin, this is moreover the case since for the kinetic theory an axial vector distribution was required, whereas for the index theorem no such introduction by hand would be necessary. Even though adiabaticity mars calculation of the Dirac index, by application of the Fujikawa method for BZ spinors, we find a chiral rotation shares some similarities to the similarity transform leading to a non-covariant Berry phase.

To achieve the classical kinetic theories, new approaches using the Berry phase on the phase space worldline were found which merit future study. Notably, we have only examined the adiabatic case, however a non-adiabatic limit, i.e., with $2pT\ll 2\pi \hbar$, requires further investigation. This is important because this well incorporates both the UV limit in propertime and, in the on-shell limit, massless fermions. Also, a non-commutative structure is present in the Berry transformed worldline Hamiltonian that may point to new physics and will be studied in another work.

\begin{acknowledgments}
We would like to thank Kenji Fukushima, Di-Lun Yang, and Pablo Morales with whom fruitful discussions have led to the betterment of this study. In particular we would like to acknowledge Kenji Fukushima for his initial involvement and for proposing the idea of a worldline kinetic theory.  S.P.  is supported by National Natural Science Foundation of China (NSFC) under Grants No. 12075235 and 12135011.
\end{acknowledgments}

\bibliography{references,qkt-ref}
\bibliographystyle{apsrev4-1}

\end{document}